\documentclass [prd,nofootinbib]  {revtex4}
\usepackage{amssymb}
\usepackage{amsmath}
\usepackage{bm}% bold math
\usepackage{graphicx}% Include figure files
\begin{document}

\title{%
A bound on Planck-scale modifications of the energy-momentum composition rule from atomic interferometry}
\author{Michele Arzano}
\email{marzano@uu.nl} \affiliation{%
Institute for Theoretical Physics and Spinoza Institute,\\
Utrecht University, Leuvenlaan 4,\\
Utrecht 3584 TD, The Netherlands}
\author{Jerzy Kowalski-Glikman}
\email{jkowalskiglikman@ift.uni.wroc.pl} \affiliation{Institute for
Theoretical Physics,\\ University of Wroc\l{}aw, Pl.\ Maxa Borna 9,\\
Pl--50-204 Wroc\l{}aw, Poland}
\author{Adrian Walkus}
\email{walkus@ift.uni.wroc.pl} \affiliation{Institute for Theoretical Physics,\\
University of Wroc\l{}aw, Pl.\ Maxa Borna 9,\\ Pl--50-204 Wroc\l{}aw, Poland}

\date{\today}

\begin{abstract}
High sensitivity measurements in atomic spectroscopy were recently used in \cite{AmelinoCamelia:2009am} to constraint the form of possible modifications of the energy-momentum dispersion relation resulting from Lorentz invariance violation (LIV).  In this letter we show that the same data can be used successfully to set experimental bounds on deformations of the energy-momentum composition rule.  Such modifications are natural in models of deformed Lorentz symmetry which are relevant in certain quantum gravity scenarios.  We find the bound for the deformation parameter $\kappa$ to be a few orders of magnitude below the Planck scale and of the same magnitude as the next-to-leading order effect found in \cite{AmelinoCamelia:2009am}. We briefly discuss how it would be possible to distinguish between these two scenarios.
\end{abstract}

\maketitle

The question of the validity of Lorentz invariance (LI) is one of the major issues in quantum gravity.  Indeed it is generally expected that in regimes where quantum and gravitational effects play an important role the very concept of space-time as a manifold should break down. This situation is expected to occur when probing the space-time structure with a resolution of the order of the Planck length  $L_p= 10^{-35}\, m$.  At this scale various quantum gravity scenarios suggest that space-time should exhibit some kind of discreteness or 'granularity' and this in turn would affect the fate of Lorentz invariance (see \cite{AmelinoCamelia:2002fw} for an in depth discussion).  In this context the most studied scenario (for recent reviews see \cite{Mattingly:2005re}, \cite{Jacobson:2005bg}, and also \cite{AmelinoCamelia:2008qg}) is that of Lorentz invariance violation (LIV).  Such departures from Lorentz invariance would manifest via modifications of the energy-momentum dispersion relation for a particle which, to the leading order in Planck mass $M_P\sim 1/L_P \sim 10^{19}\, GeV$, have the form
 \begin{equation}\label{01}
    E = \sqrt{m^2+\mathbf{p}^2} + \frac\zeta{M_P}\,  \mathbf{p}^2+O(1/M^2_P)\, ,
\end{equation}
where $\zeta$ is a numerical parameter which can be, in principle, determined from the underlying fundamental theory. Another type of LIV with the leading order term of the form $\sim m |\mathbf{p}|/M_P$ was considered in \cite{Alfaro:2001rb}.  This deviation from the ordinary dispersion relation is somewhat analogous to what happens in condensed matter systems where the dispersion relation of a propagating wave gets modified when its wave-length becomes of the order of the fundamental discreteness scale of the system.

Another possibility is that the concept of Lorentz invariance itself becomes ``deformed" as we approach the Planck scale. In this case the action of Lorentz symmetry generators on states and operators of the quantum systems under consideration becomes non linear and, most importantly, non-additive on states describing composite systems.  These type of deformations introduce an energy scale, the deformation parameter, which is invariant under the new symmetry transformations.  The main motivation for studying these departures from LI is that they arise as symmetries of the phase space of particles in three dimensional gravity whose quantization is rather well understood \cite{Freidel:2005bb}, \cite{Freidel:2005me}, \cite{Noui:2006ku}, \cite{Meusburger:2003hc}.  The effects one can derive in this case, as we discuss in detail below, are again suppressed by ratios of the energies involved in the processes and the Planck energy.

Given the extremely high energy scale at which quantum gravity effects should become manifest (the Planck energy $10^{19}\, GeV$) it was commonly believed that we could not have any hope, in a reasonable future, to experimentally observe effects that could lead our search of a unified theory of quantum and gravitational phenomena.  Therefore it came quite as a surprise when it was realized that indeed there were some experimental settings in which one could collect data to constraint possible deformations of the dispersion relation discussed above.  As a consequence over the last ten years a new research line in ``quantum -gravity phenomenology" has emerged \cite{AmelinoCamelia:1999zc}, \cite{AmelinoCamelia:2003ex}, \cite{AmelinoCamelia:2008qg}, and recently the first experimental data concerning LIV induced by quantum gravity have been put under scrutiny \cite{AmelinoCamelia:2009pg}.

In general the experimental search for new physics proceeds along two different frontiers which are somehow complementary: the high-energy one and the precision measurement one.  The phenomenological contexts in which one seeks signatures of LIV encoded in a modified dispersion relation of the type (\ref{01}) are usually associated with high-energy phenomena involving cosmic rays and gamma rays (see \cite{AmelinoCamelia:2008qg} and references therein).  In these cases the leading order quantum gravity corrections to the special relativistic dispersion relation reach an order of magnitude which falls within present or near future experimental sensitivity.  Recently, however, also precision measurements in atom interferometry have been used to put constraints on LIV \cite{AmelinoCamelia:2009am}. The authors of this paper exploited the exceptional sensitivity of atom-recoil experiments \cite{Ram1}, \cite{Ram2}, \cite{Ram3} combined with the high precision measurements of the fine structure constant \cite{gab08} to set constraints to the parameters of the non-relativistic limit of (\ref{01})
\begin{equation}\label{nonrelDR}
E\simeq m + \frac{{\bf p}^2}{2m}+\frac{1}{2M_P}\left(\xi_1 m |{\bf p}| + \xi_2 {\bf p}^2 \right)\, .
\end{equation}

We show here that the same high precision data can be used to {\it constrain scenarios with deformed Lorentz invariance}.  In our case the corrections one should look for are solely due to the non-trivial way in which momenta of scattering particles combine in a deformed symmetry setting. As we will see the deformed momentum composition rules will lead to corrections which are comparable in magnitude to the correction obtained from the $\xi_2$ term in (\ref{nonrelDR}) found in \cite{AmelinoCamelia:2009am}.  If such an effect would be observed, comparison with data from high energy observations would enable one to discriminate whether it is coming from one scenario or the other.

We start by briefly recalling how the Raman effect is used in atomic interferometry to determine the fine structure constant.  The process we are interested in is as follows. A non-relativistic (cold) atom of rest mass $m$ ($c=1$ in our convention) and momentum $p$ ($p\ll m$) absorbs a photon of energy $h\nu$ and then undergoes a stimulated emission, in opposite direction of a photon of energy $h\nu'$ with the final atom momentum equal $p_f$.  From conservation of energy one finds
\begin{equation}\label{energ}
h\Delta \nu=\frac{p_f^2-p^2}{2m}
\end{equation}
where $\Delta \nu=\nu-\nu'$.  The final momentum $p_f$ can be expressed in terms of the photon frequencies and initial momentum $p$ using momentum conservation
\begin{equation}
p_f = p + h (\nu+\nu')\, ,
\end{equation}
which can be rewritten in terms of a resonant frequency of the atom $\nu_*$ as
\begin{equation}\label{mome}
p_f = p + 2 h \nu_*\, .
\end{equation}
Substituting (\ref{mome}) in (\ref{energ}) one finds
\begin{equation}
\frac{\Delta \nu}{2\nu_*(\nu_*+p/h)}=\frac h m
\end{equation}
Measurements of $\Delta \nu/2\nu_*(\nu_*+p/h)$ allow to determine with extreme precision the ratio $\frac h m$ for the atom under consideration and such value can be used to determine the square of the fine structure constant $\alpha^2=2 R_{\infty}\frac m m_{e} \frac h m$ where $m_e$ is the electron mass and $R_{\infty}$ is the Rydberg constant \cite{Ram1}, \cite{Ram2}, \cite{Ram3}, \cite{gab08}.

Note that in the process we have a hierarchy of energy scales, with the rest mass of the atom $m \sim 10^2\, GeV$ being much larger than $p, p_f$ (the atom is moving non-relativistically with the velocity of few $cm/s$ \cite{Ram1}) and $h\nu$, $h\nu'$ which are of order of few $eV$.  We consider quantum gravity corrections to the process which will be characterized by the deformation scale $\kappa$ which is expected to be of order of the Planck mass,  $\kappa = \eta\, M_P$, with $\eta$ being the parameter, whose exact value can be, in principle read off from a fundamental theory.

%\newline
Motivations to study deformations of Lorentz symmetry come from various quantum gravity contexts.  The clearest link between relativistic symmetry deformation and quantum gravity emerges in the context of three-dimensional quantum gravity.  In this case several works have shown that particles coupled to (quantum) gravity are described by (representations of) a deformation of the three dimensional Poincar\'e algebra  \cite{Freidel:2005bb}, \cite{Freidel:2005me}, \cite{Noui:2006ku}, \cite{Meusburger:2003hc}.  Translation generators belonging to such algebra act on multi-particle states via a modification of Leibniz rule governed by a deformation parameter given by the Planck mass.  In the four-dimensional context our understanding of quantum gravity is much poorer but various arguments \cite{AmelinoCamelia:2003xp}, \cite{Freidel:2003sp}, \cite{Arzano:2007nx} suggest that a similar deformation of the Poincar\'e algebra, known as the $\kappa$-Poincar\'e algebra \cite{Lukierski:1991pn}, \cite{Lukierski:1991ff}, \cite{Majid:1994cy} might emerge in the no-gravity limit of a yet-to-be discovered quantum gravity theory. The mathematical scheme of $\kappa$--Poincar\'e algebra has been one of the inspirations of the model of relativistic kinematics with two observer-independent scales put forward in the seminal papers \cite{Amelino-Camelia:2000ge}, \cite{Amelino-Camelia:2000mn}. For a detailed account of the $\kappa$-Poincar\'e algebra and applications we refer the reader to \cite{KowalskiGlikman:2004qa}. 

This is the model we will consider to study possible quantum gravity corrections to the process described above.   Here it suffices to discuss how the deformation affects the kinematics of the atom-recoil process.  In contrast to the LIV framework in our case we have a deformation of the energy-momentum conservation condition while the energy-momentum dispersion relation remains unaltered (see \cite{Freidel:2006gc}, \cite{Freidel:2007hk}), i.e., for all constituents of the process the relativistic mass shell condition of the standard form $E^2=p^2+m^2$ holds. As for the deformed energy momentum conservation rule we use the results of \cite{KowalskiGlikman:2009zu}, where these rules were derived from the expression for the $\phi^4$ interaction term, calculated in the star product formalism developed in \cite{Freidel:2006gc}, \cite{Freidel:2007hk}.
In the case of the process $A+h\nu \rightarrow A_f + h\nu'$, where $A,A_f$ denotes the state of the atom before and after the process, in the leading order in inverse $\kappa$  the energy conservation  has the form
\begin{equation}\label{1}
    E_A+h\nu+\frac{E_A\, h\nu}\kappa=E_{A_f}+h\nu'+\frac{E_{A_f}\, h\nu'}\kappa\, .
\end{equation}
Note that this expression is symmetric in the pairs of incoming and outgoing constituents of the process, $E_A, h\nu$ and  $E_{A_f}, h\nu'$ on the left and the right hand side, respectively. Since $E_A = m + p^2/2m$, $p\ll m$, with a similar expression holding for $E_{A_f}$, in the deformation terms one can replace $E_A$ and $E_{A_f}$ with $m$. Notice that although $m$ is present in the expressions for $E_A$ and $E_{A_f}$ in the first terms of the left and the right hand sides it cancels out. Thus, neglecting sub-leading terms, we find the final expression for energy conservation to be
\begin{equation}\label{2}
   h \Delta\nu\left(1 + \frac m\kappa\right)=\frac{p_f^2-p^2}{2m }
\end{equation}
with $\Delta \nu=\nu-\nu'$. In the case of momentum conservation the deformed expression is not symmetric and therefore we deal not with one but with four independent conservation rule.  These represent different ``channels" for our process which are equally probable \cite{AmelinoCamelia:2001fd}, \cite{KowalskiGlikman:2009zu} (see the discussion below.) However since the leading order correction in (\ref{2}) is of order $m/\kappa$ in the momentum conservation rules we can safely neglect the terms of the form $p^2/2m\kappa$ or ${h\Delta \nu}/{\kappa}$ which are much smaller. As a result, in the leading order we find only three channels, to wit
\begin{eqnarray}
p+h\nu& = & -h\nu' + p_f \label{c1}\\
h\nu+ p\left(1 - \frac m\kappa\right)& = &-h\nu' + p_f\left(1-\frac{m}\kappa\right)\label{c2}\\
p+h\nu\left(1 + \frac m\kappa \right)& = &p_f-h\nu' \left(1+\frac{m}\kappa \right)\label{c3}\, .
\end{eqnarray}
Note that (\ref{c1}) is just the undeformed momentum conservation, while (\ref{c2}) and (\ref{c3}) are identical to the leading order. Thus we have to do with only two distinct channels.

To find our final expression we have to find $p_f$ for both cases and substitute it to (\ref{2}). The final result for (\ref{c1}) is
\begin{equation}\label{9}
    \frac{\Delta\nu}{2\nu_*(\nu_*+p/h)}\left[1+\frac m\kappa \right] = \frac hm
\end{equation}
while for (\ref{c2}) and (\ref{c3}) one has
\begin{equation}\label{10}
    \frac{\Delta\nu}{2\nu_*(\nu_*+p/h)}\left[1+\frac m\kappa\left(1-\frac{h\nu_*}{h\nu_* +p}\right)\right] = \frac hm
\end{equation}
We see that in both cases the correction is of the order of $m/\kappa$ (note that ${h\nu_*}/({h\nu_*+p}) \sim 10^{-2}$ for the typical experiment \cite{Ram1}, \cite{Ram2}, \cite{Ram3}.) On the other hand, since the accuracy of the classical formula (with no $\kappa$ correction) has been measured to the order of one part in $10^9$ we can conclude that $\kappa > 10^{11}\, GeV$. This bound is still eight orders of magnitude below the Planck scale but given the great progress of cold atom experimental techniques in the last years \cite{Cadoret:2008st} we can look optimistically towards tighter constraints on the value of $\kappa$ in the near future.

The results (\ref{9}) and (\ref{10}) are the same order of magnitude as in the case of quadratically deformed dispersion relation in the Lorentz breaking scheme reported in \cite{AmelinoCamelia:2009am}, and thus if the effect is observed these two schemes seem at the first sight hardly distinguishable. However, combining this possible observations with the outcomes of other experiments will make it possible to easily distinguish them. It is crucial to recall at this point that, contrary to what happens in our scenario with deformed Lorentz invariance, LIV schemes lead to a general prediction of energy dependence of the speed of light or {\it in vacuo} dispersion. This effect, although disfavored by the FERMI satellite measurements \cite{Collaborations:2009zq}, is still not ruled out \cite{AmelinoCamelia:2009pg}. If the energy dependence of the speed of light is indeed observed our model of deformed Lorentz symmetry is presumably ruled out, the same holding for LIV models if the effect is shown to be not present.

Apart from these considerations there is one more important distinction between LIV and deformed Lorentz symmetry models based on deformation of the energy-momentum composition rule.  As we saw above the latter predicts that instead of the usual single energy-momentum conservation there are many conservation channels, each occurring with equal probability \cite{AmelinoCamelia:2001fd}. This is the major novel feature of the deformed composition rule, which, if present, should be seen in sufficiently high precision measurement. However in the present context this would require to raise the sensitivity of the class of experiments we considered by ten orders of magnitude to be able to distinguish between (\ref{9}) and (\ref{10}).  Nevertheless the fact that it is possible to establish a bound on the symmetry deformation parameter $\kappa$ using precision measurements in low energy processes raises hopes that the characteristic features of deformed Lorentz invariance might lead to observable effects in  future experiments.

\begin{acknowledgments}
We thank Giovanni Amelino-Camelia for his useful comments on the first version of this paper.

The work of MA is supported by a Marie Curie Intra-European Fellowship. MA would also like to thank the University of Wroc\l{}aw for hospitality and the ``Quantum Geometry and Quantum Gravity" programme of the European Science Foundation for financial support through a Short Visit Grant.  JKG is supported in part by research projects N202 081 32/1844 and NN202318534 and Polish Ministry of Science and Higher Education grant 182/N-QGG/2008/0.

\end{acknowledgments}

\end{document}